\documentclass[amsmath]{iucr}              % DO NOT DELETE THIS LINE
\usepackage{amsmath}
\usepackage{array}
\usepackage{longtable}
\usepackage{bm}
\usepackage[dvips]{graphics}
\setlongtables

\paperprodcode{a000000}      % Replace with production code if known
\paperref{xx9999}            % Replace xx9999 with reference code if known
\papertype{FA}               % Indicate type of article
\paperlang{english}          % Can be english, french, german or russian
\journalcode{A}              % Indicate the journal to which submitted
\journalyr{2003}
\journaliss{0}
\journalvol{00}
\journalfirstpage{000}
\journallastpage{000}
\journalreceived{0 XXXXXXX 0000}
\journalaccepted{0 XXXXXXX 0000}
\journalonline{0 XXXXXXX 0000}

\def\(#1){(\ref{#1})}           % Ron's equation referencing
\newcommand\rf[1]{(\ref{#1})}   % Shahar's equation referencing
\newcommand\ie{{\it i.e.\/}}

\newcommand\kv{{\textbf k}}

\newcommand\rv{{\textbf r}}

\newcommand\tv{{\textbf t}}

\renewcommand\b[1]{{\textbf b}^{(#1)}}
\newcommand\g{\gamma}
\renewcommand\d{\delta}
\newcommand\eps{\epsilon}

\newcommand\G{\Gamma}
\newcommand\sv{{\textbf S}}
\newcommand\GS{G_S}
\newcommand\GSk{G_S^{\textbf k}}
\renewcommand\= {{\equiv}}

\newcommand\cross{\!\times\!}
\newcommand\xb{{\bar x}}
\newcommand\yb{{\bar y}}
\newcommand\xyb{{\overline{xy}}}
\newcommand\zb{{\bar z}}

\newcommand\f[1]{{\Phi}_{r_#1}^{\delta}}

\newcommand\fm{{\Phi}_{m}^{\mu}}

\newcommand \bi{{\textbf b}^{(i)}}

\newcommand\feg{{\Phi}_{e}^{\gamma}}

\newcommand\rr{r_8}

\newcommand\hf{\frac{1}{2}}

\newcommand\sr[1]{#1_\zb}

\newcommand\ax[1]{#1^*2^\dagger2^{*\dagger}}

\begin{document}
\title{Magnetically-ordered quasicrystals: Enumeration of spin groups and
  calculation of magnetic selection rules}
\shorttitle{Spin groups and magnetic selection rules for quasicrystals}

% Authors' names and addresses. Use \cauthor for the main (contact) author.
% Use \author for all other authors. Use \aff for authors' affiliations.
% Use lower-case letters in square brackets to link authors to their
% affiliations; if there is only one affiliation address, remove the [a].

\cauthor{Ron}{Lifshitz}{ronlif@post.tau.ac.il}{}
\author{Shahar}{Even-Dar Mandel}

\aff{School of Physics and Astronomy, Raymond and Beverly Sackler
  Faculty of Exact Sciences, Tel Aviv University, Tel Aviv 69978,
  \country{Israel}}

% Use \shortauthor to indicate an abbreviated author list for use in
% running heads (you will need to uncomment it).

\shortauthor{Lifshitz \& Even-Dar}

\maketitle                        % DO NOT DELETE THIS LINE

\begin{synopsis}
  It is shown how to enumerate spin point groups and spin space-group
  types for quasicrystals, and use them to calculate selection rules
  for neutron diffraction experiments. Two-dimensional octagonal
  quasicrystals are considered as an example.
\end{synopsis}

\begin{abstract}
  We provide the details of the theory of magnetic symmetry in
  quasicrystals, which has previously only been outlined. We develop a
  practical formalism for the enumeration of spin point groups and
  spin space groups, and for the calculation of selection rules for
  neutron scattering experiments. We demonstrate the formalism using
  the simple, yet non-trivial, example of magnetically-ordered
  octagonal quasicrystals in two dimensions. In a companion paper
  [{\it Acta Crystallographica\/} A{\bf XX} (2003) xxx-xxx], we
  provide complete results for octagonal quasicrystals in three
  dimensions.
\end{abstract}

\section{Introduction}
\label{sec:intro}

The discovery of Zn-Mg-RE icosahedral quasicrystals, where RE stands
for one of the rare-earth metals Y, Gd, Tb, Dy, Ho, and
Er~\cite{Niikura94,Tsai94}, and initial indications that some of these
quasicrystals may possess long-range antiferromagnetic
order~\cite{Charrier97,Charrier98}, have generated increased interest
in the nature of magnetic order in quasicrystals~\cite[and references
therein]{Fukamichi99}. Subsequent measurements~\cite{Islam98,Sato98},
followed by ongoing vigorous
research~\cite{Fisher99,Sato99,Fisher00,Sato00a,Sato00b,Dolinsek01,Kramer02},
have shown that only short-range spin correlations exist in these
particular quasicrystals, and therefore the existence of long-range
magnetic order in real quasicrystals remains an unresolved question of
great interest. The recent discovery of cadmium-based
quasicrystals~\cite{Guo00a,Guo00b,Tsai00} and initial studies of
magnetic order in the Cd-Mg-Tb icosahedral quasicrystal~\cite{Sato02},
may provide some insight into this question. Theoretical models that
deal with magnetism on quasicrystals---purely geometrical
models~\cite{Niizeki90a,Niizeki90b,icq5,msea00} as well as physical
ones, such as the Ising model~\cite{Grimm97,Matsuo00,Matsuo02}, the XY
model~\cite{Hermisson00}, the Heisenberg model~\cite{wessel}, and the
Hubbard model~\cite{Jagannathan97,Hida01}---are known to exhibit
long-range magnetic order.  This is despite initial intuition that
aperiodicity necessarily induces geometrical frustration and is
therefore inconsistent with having magnetic order.

It is therefore clear---both from a theoretical and an experimental
standpoint---that there is a need for a theoretical classification of
all types of quasiperiodic long-range magnetic order that are allowed
by symmetry. Such a symmetry-based classification of quasiperiodic
magnetic structures, combined with a calculation of the selection
rules, imposed by magnetic symmetry, will be of great assistance in
the study of real magnetic quasicrystals, if and when they are
discovered. It will also offer valuable guidance in the search for
these novel materials~\cite{msea00}. In this paper we provide the
details of the extension to quasicrystals~\cite{prl98} of Litvin and
Opechowski's theory of spin groups~\cite{litvin73,litvin74,litvin77}.
In section~\ref{sec:groups} we explain how spin groups are used to
describe the magnetic symmetry of periodic as well as quasiperiodic
crystals. In section~\ref{sec:classification} we give the group
theoretic details of the classification of the different groups into
appropriate equivalence classes. In section~\ref{sec:enumeration} we
develop the formalism required for the actual enumeration of spin
groups, and in section~\ref{sec:selection} we derive the relations
between the magnetic symmetry of a crystal and the selection rules it
imposes on neutron diffraction experiments. In section~\ref{sec:octa}
we treat the case of octagonal symmetry in two dimensions as a
pedagogical example. In a companion paper~\cite{octa3d} we perform the
actual detailed enumeration of three-dimensional octagonal spin groups
which turns out to be surprisingly rich. Complete enumeration for the
other common quasicrystals without explicit details will follow in
future publications.

\section{Using spin groups to describe the symmetry of
  magnetically-ordered crystals}
\label{sec:groups}

A $d$-dimensional magnetically-ordered crystal, whether periodic or
aperiodic, is most directly described by its spin density field
$\sv(\rv)$. This field is a 3-component real-valued function,
transforming like an axial vector under $O(3)$ and changing sign under
time inversion.  One may think of this function as defining a set of
classical magnetic moments, or spins, on the atomic sites of the
material.\footnote{We usually consider 3-dimensional magnetic moments,
  or spins, in a $d$-dimensional crystal, where $d=2$ or 3, and
  therefore take $\sv(\rv)$ to be a 3-component field. If necessary,
  one can generalize to spins of arbitrary dimension.}  For
quasiperiodic crystals the spin density field may be expressed as a
Fourier sum with a countable infinity of wave vectors
\begin{equation}\label{FourierS}
\sv(\rv)=\sum_{\kv\in L} \sv(\kv) e^{i\kv\cdot\rv} .
\end{equation}
The set $L$ of all integral linear combinations of the wave vectors in
\(FourierS) is called the {\it magnetic lattice,} and is characterized
among other things by a rank $D$, and by a lattice point group $G_L$:
Its {\it rank\/} $D$ is the smallest number of wave vectors needed to
generate it by integral linear combinations.  For quasiperiodic
crystals, by definition, the rank is finite. For the special case of
periodic crystals the rank is equal to the dimension $d$ of physical
space. The set of (proper or improper) rotations, that when applied to
the origin of Fourier space leave the magnetic lattice invariant, is
the {\it lattice point group\/} $G_L$, also called the {\it
  holohedry.}

The theory of magnetic symmetry in quasiperiodic crystals
\cite{prl98}, is a reformulation of Litvin and Opechowski's theory of
spin space groups~\cite{litvin73,litvin74,litvin77}.  Their theory,
which is applicable to periodic crystals, is extended to quasiperiodic
crystals by following the ideas of Rokhsar, Wright, and Mermin's
``Fourier-space approach'' to
crystallography~\cite{rwm2,rwm1}.\footnote{For a review, see
  \citeasnoun{merminreview} or \citeasnoun{msota}, for an elementary
  introduction, see \citeasnoun{physa96}.} At the heart of this
approach is a redefinition of the concept of point-group symmetry
which enables one to treat quasicrystals directly in physical space,
as opposed to the alternative ``superspace approach'' \cite{volc}.
The key to this redefinition is the observation that point-group
rotations (proper or improper), when applied to a quasiperiodic
crystal, do not leave the crystal invariant but rather take it into
one that contains the same spatial distributions of bounded structures
of arbitrary size.

This generalized notion of symmetry, termed {\it
  indistinguishability,} is captured by requiring that any symmetry
operation of the magnetic crystal leave invariant all
spatially-averaged autocorrelation functions of its spin density field
$\sv(\rv)$, for any order $n$ and for any choice of components
$\alpha_i\in \{x,y,z\}$,
\begin{equation}\label{corr2}
 C^{(n)}_{\alpha_1\ldots\alpha_n}(\rv_1,\ldots,\rv_n)
 =\lim_{V\to\infty}\frac1{V}\int_V d\rv
   S_{\alpha_1}(\rv_1-\rv)\cdots S_{\alpha_n}(\rv_n-\rv).
\end{equation}

It has been shown \cite[in the Appendix]{rmp97} that an equivalent
statement for the indistinguishability of any two quasiperiodic spin
density fields, $\sv(\rv)$ and $\sv'(\rv)$, is that their Fourier
coefficients are related by
\begin{equation}\label{chidef}
\sv'(\kv) = e^{2\pi i\chi(\kv)}\sv(\kv),
\end{equation}
where $\chi$, called a {\it gauge function,} is a real-valued scalar
function which is linear (modulo integers) on the magnetic lattice
$L$. This simply means that
\begin{equation}
  \label{eq:chilin}
  \forall \kv_1,\kv_2\in L:\quad \chi(\kv_1+\kv_2) \= \chi(\kv_1) +
  \chi(\kv_2),
\end{equation}
where ``$\=$'' denotes equality modulo integers.

With this in mind, we define the {\it point group $G$\/} of a
$d$-dimensional magnetic crystal to be the set of operations $g$ in
$O(d)$ that leave it indistinguishable to within rotations $\g$ in
spin space, possibly combined with time inversion.\footnote{Note that
  since $\sv(\rv)$ is an axial vector field we can restrict $\g$ to be
  a proper rotation without any loss of generality.} Accordingly, for
every pair $(g,\g)$ there exists a gauge function, $\Phi_g^\g(\kv)$,
called a {\it phase function}, which satisfies
\begin{equation}\label{phase}
\sv(g\kv) = e^{2\pi i \Phi_g^\g(\kv)}\g\sv(\kv).
\end{equation}
In general, as we shall see later, there may be many spin-space
operations $\g$ that, when combined with a given physical-space
rotation $g$, satisfy the point-group condition \(phase). We denote
physical-space rotations by Latin letters and spin-space operations by
Greek letters. We use a primed Greek letter to explicitly denote the
fact that a spin-space rotation is followed by time inversion. Thus,
the identity rotation in physical-space is $e$, the identity rotation
in spin-space is $\eps$, and time inversion is $\eps'$. Also note that
we use the same symbol $\g$ to denote an abstract spin-space operation,
and to denote the $3\cross 3$ matrix, operating on the field $\sv$,
representing this operation.

If $(g,\g)$ and $(h,\eta)$ both satisfy the point group condition
\(phase), then it follows from the equality
\begin{equation}
  \label{eq:product}
  \sv([gh]\kv)=\sv(g[h\kv]), 
\end{equation}
that so does $(gh,\g\eta)$. This establishes that the set $\G$ of all
transformations $\g$ forms a group, and the set $\GS$ of all pairs
$(g,\g)$, satisfying the point-group condition \(phase), also forms a
group. The latter is a subgroup of $G\times\G$, called the {\it spin
  point group\/}. We shall consider here only finite groups $G$ and
$\G$, although in general this need not be the case.  The equality
\(eq:product) further implies that the corresponding phase functions,
one for each pair in $\GS$, must satisfy the {\it group compatibility
  condition,}
\begin{equation}\label{GCC}
\forall (g,\g), (h,\eta)\in\GS:\ \Phi_{gh}^{\g\eta}(\kv) \=
\Phi_g^\g(h\kv) + \Phi_h^\eta(\kv).
\end{equation}
Note that successive application of the group compatibility condition
\(GCC) reveales a relatively simple relation between the phase
functions of two conjugate elements $(g,\g)$ and
$(hgh^{-1},\eta\g\eta^{-1})$ of $\GS$,
\begin{equation}\label{eq:conj}
\forall (g,\g), (h,\eta)\in\GS:\
\Phi_{hgh^{-1}}^{\eta\g\eta^{-1}}(h\kv) \= \Phi_g^{\g}(\kv) +
\Phi_h^{\eta}(g\kv-\kv).
\end{equation}

A {\it spin space-group,} describing the symmetry of a magnetic
crystal, whether periodic or aperiodic, is thus given by a magnetic
lattice $L$, a spin point group $\GS$, and a set of phase functions
$\Phi_g^\g(\kv)$, satisfying the group compatibility condition \(GCC).
We continue to call this a spin space-group even though its
physical-space part is no longer a subgroup of the Euclidean group
$E(d)$.  Nevertheless, the spin space-group may be given an algebraic
structure of a group of ordered triplets $(g,\g,\Phi_g^\g)$ in a
manner similar to the one shown originally by~\citeasnoun{rhm}, and
more recently again by~\citeasnoun{dm}, in the context of ordinary
space groups for nonmagnetic crystals.

In the case of periodic crystals one can show
\cite[Eq.~(2.18)]{merminreview} that any gauge function
$2\pi\chi(\kv)$, relating two indistinguishable spin density fields as
in Eq.~\rf{chidef}, is necessarily of the form \hbox{$\kv\cdot\tv$} for
some constant translation vector $\tv$ independent of $\kv$, so that
$\sv'(\rv)=\sv(\rv+\tv)$ and indistinguishability reduces to identity
to within a translation.  One can then combine rotations in physical
space and in spin space with translations to recover the traditional
spin space-groups of periodic crystals, containing operations that
satisfy
\begin{equation}\label{real}
\sv(g\rv)=\g\sv(\rv+\tv_g^\g)\ ,
\end{equation}
leaving the spin density field {\it identical\/} to what it was. In
the quasiperiodic case one must retain the general form of
$\Phi_g^\g(\kv)$ which is defined only on the magnetic lattice and
cannot be linearly extended to arbitrary~$\kv$.

\section{Classification of spin groups}
\label{sec:classification}

The common symmetry properties of different magnetic structures become
clear only after they are classified into properly chosen equivalence
classes. We are concerned here with the classification of magnetic
crystals into Bravais classes (section~\ref{sec:bravais}), spin
geometric crystal classes (section~\ref{sec:geometric}), spin
arithmetic crystal classes (section~\ref{sec:arithmetic}), and spin
space-group types (section~\ref{sec:types}).

\subsection{Bravais classes}
\label{sec:bravais}

Magnetic crystals, as well as nonmagnetic crystals, are classified
into Bravais classes according to their lattices of wave vectors.
Intuitively, two magnetic lattices are in the same Bravais class if
they have the same rank $D$ and point group $G_L$ (to within a spatial
reorientation) and if one can `interpolate' between them with a
sequence of lattices, all with the same point group and rank.  Stated
more formally, as presented by~\citeasnoun{dm}, we say that two
magnetic lattices $L$ and $L'$ belong to the same {\it Bravais
  class\/} if:
\begin{enumerate}
\item The two lattices are isomorphic as abelian groups, \ie\ there is
a one to one mapping, denoted by a prime ($'$), from $L$ onto $L'$
\begin{equation}\label{prime}
\begin{array}{rccc}
': & L & \longrightarrow & L'  \\
   & \kv & \longrightarrow & \kv' \\
\end{array}
\end{equation}
%that takes every wave vector $\kv\in L$ into a wave vector $\kv'\in L'$
satisfying
\begin{equation}\label{bravais1}
(\kv_1 + \kv_2)' = \kv_1' + \kv_2'\ ;
\end{equation}
\item The corresponding lattice point groups $G_L$ and $G_L'$ are
conjugate subgroups of $O(d)$,
\begin{equation}\label{bravais2}
G_L' = rG_Lr^{-1}\ ,
\end{equation}
for some proper $d$-dimensional rotation $r$; and
\item The isomorphism \(prime) between the lattices preserves the
actions of their point groups, namely
\begin{equation}\label{bravais3}
(g\kv)' = g'\kv'\ ,
\end{equation}
where $g'=rgr^{-1}$.
\end{enumerate}
Since the classification of magnetic lattices for magnetic crystals is
the same as the classification of ordinary lattices for nomagnetic
crystals we shall not expand on this issue further but rather refer
the interested reader to previous discussions on the
matter~\cite{rmw,mrw,mrrw,merminreview,ml,merminprl,physa96,dm}.

\subsection{Spin geometric crystal classes}
\label{sec:geometric}

When we say that two magnetic crystals ``have the same spin point
group'' we normally mean that they belong to the same equivalence
class of spin point groups, called a spin geometric crystal class.  We
say that two spin point groups $\GS$ and $\GS'$ are in the same {\it
  spin geometric crystal class\/} if they are conjugate subgroups of
$O(d)\cross [SO(3)\cross 1']$, where $1'$ is the time inversion group,
containing the identity $\eps$ and the time inversion operation
$\eps'$. This simply means that
\begin{equation}\label{geometric}
\GS' = (r,\sigma)\GS(r,\sigma)^{-1}\ ,
\end{equation} for some physical-space rotation $r\in O(d)$, and some
spin-space operation $\sigma\in SO(3)\cross1'$.  The effect of these
rotations on the spin point group $G_S$ is to reorient its symmetry
axes both in physical space and in spin space.

\subsection{Spin arithmetic crystal classes}
\label{sec:arithmetic}

The concept of a spin arithmetic crystal class is used to distinguish
between magnetic crystals which have equivalent magnetic lattices and
equivalent spin point groups but differ in the manner in which the
lattice and the spin point group are combined. Two magnetic crystals
belong to the same {\it spin arithmetic crystal class\/} if their
magnetic lattices are in the same Bravais class, their spin point
groups are in the same spin geometric crystal class, and it is
possible to choose the lattice isomorphism \(prime) such that the
proper rotation $r$ used in \(bravais2) to establish the lattice
equivalence is the same rotation used in \(geometric) to establish the
spin point group equivalence.

\subsection{Spin space-group types}
\label{sec:types}

The finer classification of crystals in a given spin arithmetic
crystal class into {\it spin space-group types} is an organization of
sets of phase functions into equivalence classes according to two
criteria:

\begin{enumerate}
\item Two {\it indistinguishable\/} magnetic crystals $\sv$ and
  $\sv'$, related as in \(chidef) by a gauge function $\chi$, should
  clearly belong to the same spin space-group type. Such crystals are
  necessarily in the same spin arithmetic crystal class but the sets
  of phase functions $\Phi$ and $\Phi'$ used to describe their space
  groups may, in general, be different. It follows directly from
  \(chidef) and from the point group condition \(phase) that two such
  sets of phase functions are related by
  \begin{equation}\label{gauge-tr}
  {\Phi'}_{g}^{\g}(\kv) \= \Phi_{g}^{\g}(\kv) + \chi(g\kv-\kv)\ ,
  \end{equation}
  for every $(g,\g)$ in the spin point group and every $\kv$ in the
  magnetic lattice. We call two sets of phase functions that describe
  indistinguishable spin density fields {\it gauge-equivalent\/} and
  equation \(gauge-tr), converting $\Phi$ into $\Phi'$, a {\it gauge
    transformation}. The freedom to choose a gauge $\chi$ by which to
  transform the Fourier coefficients $\sv(\kv)$ of the spin density
  field and all the phase functions $\Phi$, describing a given
  magnetic crystal, is associated in the case of periodic magnetic
  crystals with the freedom one has in choosing the real-space origin
  about which all the point group operations are applied.

\item Two {\it distinguishable\/} magnetic crystals $\sv$ and $\sv'$,
  whose spin space-groups are given by magnetic lattices $L$ and $L'$,
  spin point groups $\GS$ and $\GS'$, and sets of phase functions
  $\Phi$ and $\Phi'$, have the same spin space-group type if they are
  in the same spin arithmetic crystal class, and if, to within a
  gauge transformation \(gauge-tr), the lattice isomorphism \(prime)
  taking every $\kv\in L$ into a $\kv'\in L'$ preserves the values of
  all the phase functions
  \begin{equation}\label{scale-tr}
  {\Phi'}_{g'}^{\g'}(\kv') \= \Phi_{g}^{\g}(\kv)\ ,
  \end{equation}
  where $g'=rgr^{-1}$ and $\g'=\sigma\g\sigma^{-1}$. Two sets of phase
  functions that are related in this way are called {\it
    scale-equivalent\/}. This nomenclature reflects the fact that the
  lattice isomorphism \(prime) used to relate the two magnetic
  lattices may often be achieved by rescaling the wave vectors of one
  lattice into those of the other.
\end{enumerate}

\section{Enumeration of spin groups}
\label{sec:enumeration}

The task of enumerating spin groups is limited to the enumeration of
the distinct types of spin point groups and spin space-groups. This is
because the classification of magnetic lattices into Bravais classes,
as well as the determination of all distinct relative orientations of
point groups $G$ with respect to these lattices, giving rise to
different arithmetic crystal classes, are the same as for nonmagnetic
crystals, and therefore need not be enumerated again. The enumeration
of possible spin point groups and spin space-groups is greatly
simplified if one first lists all the general conatraints these groups
must obey due to their algebraic structure. We list below the general
constraints on the spin point group $\GS$ (section~\ref{sec:cosets}),
discuss the consequences of these constraints on the group of
spin-space operations $\G$ (section~\ref{sec:gamma}), describe a
particularly interesting connection between a certain subgroup of $\G$
and the magnetic lattice $L$ (section~\ref{sec:iso}), and then outline
the sequence of steps taken in the enumeration of spin groups
(section~\ref{sec:flow}).

\subsection{Structure of the spin point group $G_S$}
\label{sec:cosets}

The algebraic structure of the spin point group $G_S$ is severely
constrained by the point group condition \(phase) as described by the
five statements below. Proofs for the first four statements can be
found in the review on color symmetry \cite[Section IV.A.]{rmp97} as
they apply equally to the structure of the {\it color\/} point group
of a colored crystal.

\begin{enumerate}
\item The set of real-space operations associated with
  the spin-space identity $\eps$ forms a normal subgroup of $G$,
  called $G_\eps$.

  Note that as a special case of Eq.~\rf{eq:conj} the phase functions
  of conjugate elements of $G_\eps$ are related by
  \begin{equation}\label{gccGeps1}
  \forall g\in G_\eps,(h,\eta)\in G_S:\quad \Phi_{hgh^{-1}}^\eps(h\kv) \=
  \Phi_g^\eps(\kv) + \Phi_h^\eta(g\kv-\kv).
  \end{equation}
\item The set of spin-space operations paired with the real-space
  identity $e$ forms a normal subgroup of $\G$, called the {\it
    lattice spin group\/} $\G_e$. In the special case of periodic
  crystals, the elements of $\G_e$ are spin-space operations that,
  when combined with translations, leave the magnetic crystal
  invariant.

  Again, as a special case of Eq.~\rf{eq:conj} the phase functions
  of conjugate elements of $\G_e$ are related by
  \begin{equation}\label{eq:gccGe}
  \forall \g\in\G_e,(h,\eta)\in G_S:\quad \Phi_e^{\eta\g\eta^{-1}}(h\kv) \=
  \Phi_e^\g(\kv).
  \end{equation}
\item The lattice spin group $\G_e$ is abelian.
\item The pairs in $G_S$ associate all the elements of each coset of
  $G_\eps$ with all the elements of a single corresponding coset of
  $\G_e$. This correspondence between cosets is an isomorphism between
  the quotient groups $G/G_\eps$ and $\G/\G_e$.
\item If two phase functions $\Phi_e^{\g_1}(\kv)$ and
  $\Phi_e^{\g_2}(\kv)$, associated with the lattice spin group $\G_e$,
  are identical on all wave vectors then $\g_1=\g_2$.

\noindent {\sl\underline{Proof}:\/} From the point group condition
\rf{phase} we obtain
\begin{equation}\label{identical}
\forall\kv\in L:\quad \g_1\sv(\kv)=\g_2\sv(\kv),
\end{equation}
implying that the two operations have the same effect on all the spin
density fields whose symmetry is described by this paricular spin
space group type, and are therefore identical.

\end{enumerate}

\subsection{Consequences for $\G$ and $\G_e$}
\label{sec:gamma}

The lattice spin group $\G_e$ is severely constrained by being an
abelian subgroup of $SO(3)\cross 1'$. Namely, it can have no more than
a single axis of $n$-fold symmetry with $n>2$. This implies that the
possible lattice spin groups $\G_e$ are the ones listed in the first
column of Table~\ref{tab:extensions}.

The fact that the lattice spin group $\G_e$ is a normal subgroup of
$\G$ implies that $\G$ cannot contain any rotation $\sigma\in SO(3)$
for which $\sigma\G_e\sigma^{-1} \neq\G_e$. One can easily verify that
the possible supergroups $\G$ for each lattice spin group $\G_e$ are
the ones listed in the second column of Table~\ref{tab:extensions}.

\subsection{Relation between the magnetic lattice $L$ and the lattice
  spin group $\G_e$.}
\label{sec:iso}

We have already mentioned that in the special case of periodic
crystals, the lattice spin group $\G_e$ is the set of all spin-space
operations that, when combined with real-space translations,
leave the magnetic crystal invariant. It should be of no surprise then
that in the quasiperiodic case there should remain an intimate
relation between the lattice spin group $\G_e$ and the magnetic
lattice $L$. We describe this relation here without proof, which can
be found in the review on color symmetry \cite[Section IV.C.]{rmp97}
where a similar relation exists between the lattice $L$ and lattice
{\it color\/} group of a colored crystal.

Recalling that the lattice $L$ is itself an abelian group under the
addition of wave vectors, one can show that it necessarily contains a
sublattice $L_0$, invariant under the point group $G$, for which the
quotient group $L/L_0$ is isomorphic to the lattice spin group $\G_e$.
This isomorphism is established through the properties of the phase
functions $\Phi_e^\g(\kv)$ associated with all elements $\g$ of the
lattice spin group. In particular, the sublattice $L_0$ is defined as
the set of wave vectors $\kv$ for which the phases
$\Phi_e^\g(\kv)\=0$, for all elements $\g$ of the lattice spin group.
Furthermore, the relation~\(eq:gccGe) between phase functions of
conjugate elements of $\G_e$ ensures that the isomorphism between
$L/L_0$ and $\G_e$ is invariant under all elements $(h,\eta)$ of the
spin point group. In other words, if the isomorphism maps a particular
wave vector $\kv$ to a particular spin operation $\g$, then for every
$(h,\eta)$ in $\GS$, the wave vector $h\kv$ is mapped to
$\eta\g\eta^{-1}$.

This relation between the lattice spin group and the magnetic lattice
not only imposes a severe constraint on the possible lattice spin
groups but also provides an additional method to calculate the phase
functions $\Phi_e^\g(\kv)$. One of two alternative approaches can be
taken to enumerate the allowed combinations of $\G_e$ and $\G$:
\begin{enumerate}
\item For each type of lattice spin group $\G_e$, listed in
  Table~\ref{tab:extensions}, see whether there exists an invariant
  sublattice $L_0$ of $L$ giving a modular lattice $L/L_0$ isomorphic
  to $\G_e$, {\it and\/} whether the possible extensions of $\G_e$ into
  supergroups $\G$, also listed in Table~\ref{tab:extensions}, allow
  the isomorphism to be invariant under the spin point group.
\item For each type of lattice spin group $\G_e$ and its possible
  extensions into supergroups $\G$, listed in
  Table~\ref{tab:extensions}, simply try to solve all the group
  compatibilty conditions \(GCC) imposed on the phase functions
  $\Phi_e^\g(\kv)$, associated with the elements of $\G_e$ and the
  wave vectors of $L$. If a solution exists then $\G_e$ is a possible
  lattice spin group, otherwise it is not.
\end{enumerate}
It should be emphasized that, either way, the possible combinations of
$\G_e$ and $\G$, and therefore the possible types of spin point
groups, cannot be determined independently of the choice of magnetic
lattice $L$.

\subsection{Enumeration steps}
\label{sec:flow}

The enumeration of spin point groups and spin space-groups consists of
a sequence of steps which are listed schematically in the flowchart of
Fig.~\ref{fig:flow}. We shall illustrate the whole process in
section~\ref{sec:octa} by enumerating, as an example, all the
2-dimensional octagonal spin point groups and spin space-groups.

One begins by choosing a lattice $L$ from any of the known Bravais
classes. One then picks any point group $G$, compatible with $L$, and
lists all its normal subgroups $G_\eps$ along with the corresponding
quotient groups $G/G_\eps$. One then chooses one of the normal
subgroups $G_\eps$ and calculates, using one of the two approaches
described in the previous section, all allowed combinations of $\G$
and $\G_e$ such that the quotient group $\G/\G_e$ is isomorphic to
$G/G_\eps$. One then pairs the cosets of $G_\eps$ in $G$ with the
cosets of $\G_e$ in $\G$ in all distinct ways. After checking for
equivalence, as described in section~\ref{sec:geometric}, one arrives
at a list of the distinct types of spin point groups.

For each spin point group one then looks for all solutions to the
group compatibility conditions \(GCC) not already considered above.
These solutions are organized into gauge-equivalence and
scale-equivalence classes, as described in section~\ref{sec:types},
yielding the distinct spin space-group types. Because phase functions
are linear on the lattice $L$ [Eq.~\(eq:chilin)] it is sufficient to
specify their values on a chosen set of $D$ wave vectors that
primitively generate the lattice. Also, it is sufficient to specify
the phase functions only for a small set of operations $(g,\g)$ that
generate the spin point group. All other phase functions can be
determined through the group compatibility condition. Furthermore, one
can greatly simplify the calculation of phase functions by making a
judicious choice of gauge prior to solving the group compatibilty
conditions, rather than solving the group compatibility conditions and
only then organizing the solutions into gauge-equivalence classes.

\section{Calculation of magnetic selection rules}
\label{sec:selection}

Magnetic selections rules, or symmetry-imposed constraints on
the form of the spin density field, offer one of the most direct
experimental observations of the detailed magnetic symmetry of a
magnetic crystal.  In elastic neutron scattering experiments, every
wave vector $\kv$ in $L$ is a candidate for a magnetic Bragg peak,
whose intensity is given by \cite{Izyumov}
\begin{equation}\label{intensity}
I(\kv)\propto|\sv(\kv)|^2 - |\hat{\kv}\cdot\sv(\kv)|^2,
\end{equation} where $\kv$ is the scattering wave vector and
$\hat{\kv}$ is a unit vector in its direction. It has been shown
\cite{krakow} that, under generic circumstances, there can be only
three reasons for not observing a magnetic Bragg peak at $\kv$ even
though $\kv$ is in $L$: (a) The intensity $I(\kv)\neq 0$ but is too
weak to be detected in the actual experiment; (b) The intensity
$I(\kv)=0$ because $\sv(\kv)$ is parallel to $\kv$; and (c) The
intensity $I(\kv)=0$ because magnetic selection rules require the
Fourier coefficient $\sv(\kv)$ to vanish. Selection rules that lead to
a full extinction of a Bragg peak are the most dramatic and easiest to
observe experimentally. Other types of selection rules ({\it
  e.g.} that lead to an extinction of one of the components of
$\sv(\kv)$, or to a nontrivial relation between the components of
$\sv(\kv)$) are harder to observe.

We calculate the symmetry-imposed constraints on $\sv(\kv)$, for any
given wave vector $\kv\in L$, by examining all spin point-group
operations $(g,\g)$ for which $g\kv=\kv$.  These elements form a
subgroup of the spin point group which we call the {\it little spin
  group of\/} $\kv$, $\GSk$. For elements $(g,\g)$ of $\GSk$, the
point-group condition \(phase) can be rewritten as
\begin{equation}\label{eigen}
\g\sv(\kv) = e^{-2\pi i \Phi_g^\g(\kv)}\sv(\kv).
\end{equation}
This implies that every Fourier coefficient $\sv(\kv)$ is required to
be a simultaneous eigenvector of all spin-space operations $\g$ in the
little spin group of $\kv$, with the eigenvalues given by the
corresponding phase functions. If a non-trivial 3-dimensional vector
satisfying Eq.~\(eigen) does not exist then $\sv(\kv)$ will
necessarily vanish. It should be noted that the phase values in
Eq.~\(eigen) are independent of the choice of gauge~\(gauge-tr), and
are therefore uniquely determined by the spin space-group type of the
crystal.

The process of determining the form of the simultaneous eigenvector
$\sv(\kv)$ is greatly simplified if one makes the following
observation.  Due to the group compatibility condition \(GCC) the set
of eigenvalues in Eq.~\(eigen) for all the elements $(g,\g)\in\GSk$
forms a 1-dimensional representation of that group. Spin space-group
symmetry thus requires the Fourier coefficient $\sv(\kv)$ to transform
under a particular 1-dimensional representation of the spin-space
operations in the little spin group of $\kv$. We also independently
know that $\sv(\kv)$ transforms under spin-space rotations as a
3-dimensional axial vector, changing its sign under time inversion. It
is therefore enough to check whether the particular 1-dimensional
representation, dictated by the spin space-group, is contained within
the 3-dimensional axial-vector representation. If it is not, then
$\sv(\kv)$ must vanish; if it is, then $\sv(\kv)$ must lie in the
subspace of spin space transforming under this 1-dimensional
representation.

\section{Octagonal spin groups in two dimensions --- An example}
\label{sec:octa}

To demonstrate the ideas presented in this paper, we enumerate the
octagonal spin groups in two dimensions, and calculate the magnetic
selection rules that arise for each spin space-group type. We choose
to treat the octagonal crystal system because it is the most
interesting example for a magnetic quasicrystal in two-dimensions. The
reason for this is twofold: First of all, as for nonmagnetic
2-dimensional crystals, only when the order of symmetry is a power of
2 is it possible to have space groups with nonsymmorphic operations;
Secondly, only when the order of symmetry is a power of 2 is it
possible to have simple antiferromagnetic order
\cite{Niizeki90a,Niizeki90b,rmp97,msea00}.

Only partial enumerations of spin groups on quasicrystals exist to
date. Decagonal spin point groups and spin space-group types in two
dimensions have been listed by \citeasnoun{icq5} without providing
much detail regarding the enumeration process. All possible lattice
spin groups $\G_e$ for icosahedral quasicrystals have been tabulated
\cite{prl98} along with the selection rules that they impose, but a
complete enumeration of all icosahedral spin groups was not given.
This is therefore, the first complete and rigorous enumeration of
spin groups and selection rules for a quasiperiodic crystal system in
any dimension. In a companion paper \cite{octa3d} we enumerate the
octagonal spin groups in three dimensions, and in future publications
we intend to treat all the other common quasiperiodic crystal systems,
though we shall probably not include the full details of the
calculation. 

\subsection{Two-dimensional octagonal point groups and Bravais
  classes}
\label{sec:prelims}

The lowest rank $D$ that a two-dimensional octagonal lattice can have
is $4$.  There is just a single Bravais class of two-dimensional
rank-4 octagonal lattices \cite{mrw}. All lattices in this
two-dimensional Bravais class contain an 8-fold star of wave vectors
of equal length, separated by angles of $\frac{\pi}{4}$ (as shown in
Fig.~\ref{fig:star}), of which four, labeled $\b1\ldots\b4$, can be
taken as integrally-independent lattice generating vectors. The
lattice point group $G_L$ is $8mm$, generated by an eightfold rotation
$\rr$ and either a mirror of type $m$, which contains one of the
generating vectors and its negative, or a mirror of type $m'$, which
lies between two of the generating vectors. The two-dimensional point
groups $G$ to be considered in the enumeration are $8mm$ and its
subgroup $8$. There is only a single way to orient the two point groups
with respect to the lattice, so there is just a single spin arithmetic
crystal class for each spin geometric crystal class.

\subsection{Enumeration of spin point groups}

We begin by listing in the first columns of Table~\ref{tab:normal} all
normal subgroups $G_\eps$ of the point groups $G=8mm$ and $8$, and the
resulting quotient groups $G/G_\eps$. Note that the two subgroups
$2mm$ and $m$ of the point group $8mm$ are not normal and therefore do
not appear in the Table.

As generators of the spin point groups we take the generators of $G$
($\rr$ and $m$ for $G=8mm$, and $\rr$ alone for $G=8$), and combine
each one with a representative spin-space operation from the coset of
$\G_e$ with which it is paired. We denote the spin-space operation
paired with $\rr$ by $\d$ and the operation paired with $m$ by $\mu$.
When $\rr$ (or $m$) are in $G_\eps$ we take $\d$ (or $\mu$) to be
$\eps$. The constraints on the operations $\d$ and $\mu$, due to the
isomorphism between $G/G_\eps$ and $\G/\G_e$, are summarized in the
fourth column of Table \ref{tab:normal}. To the generators $(\rr,\d)$
and $(m,\mu)$ we add as many generators of the form $(e,\g_i)$ as
required, where $\g_i$ are the generators of $\G_e$ (three at the
most). Although this set of spin point-group generators may, in
general, be overcomplete, it is the most convenient set to take.

\subsection{Calculation of possible $\G$ and $\G_e$}

We use the group compatibility conditions \rf{GCC} on the phase
functions $\Phi_e^\g(\kv)$, associated with elements in the lattice
spin group $\G_e$, in order to calculate the possible combinations of
$\G$ and $\G_e$.

We first note, from inspection of Table~\ref{tab:normal}, that no
quotient group $G/G_\eps$ contains an operation of order 3. This
implies, among other things, that $\G/\G_e$ cannot contain such an
operation and therefore the extensions of the orthorhombic lattice
spin groups $\G_e$, listed in the third row of
Table~\ref{tab:extensions}, into supergroups $\G$ cannot be
cubic---they can be tetragonal at most. This then implies that for any
possible combination of $\G$ and $\G_e$,
\begin{equation}
  \label{eq:ddgdd}
  \forall \g\in\G_e, \d\in\G:\quad \d^2\g\d^{-2}=\g.
\end{equation}

With this relation at hand we can proceed to prove the following short
lemmas:
\begin{enumerate}
\item The lattice spin group $\G_e$ contains no more than 3 elements
  $\g\neq\eps$, all of which are of order 2.

\noindent {\sl\underline{Proof}:\/} Let $\d\in\G$ be the operation
paired with $\rr$ in the spin point group. The relation \rf{eq:ddgdd}
together with Eq.~\rf{eq:gccGe}, relating phase functions of conjugate
elements in $\G_e$, yield
\begin{equation}
  \label{eq:no-n}
  \Phi_e^\g(\bi)\=\Phi_e^{\d^2\g\d^{-2}}(\rr^2\bi)\=\Phi_e^\g(\rr^2\bi).
\end{equation}
Thus, for any $\g\in\G_e$
\begin{equation}
  \label{eq:phasesg}
  \Phi_e^\g(\b1) \= \Phi_e^\g(\b3) \=\alpha;\quad \Phi_e^\g(\b2) \=
  \Phi_e^\g(\b4) \=\beta;
\end{equation}
and
\begin{equation}
  \label{eq:order2}
  \Phi_e^\g(-\bi) \= \Phi_e^\g(\bi) \Longrightarrow
  \Phi_e^\g(\bi) \= 0 {\rm \ or\ } \frac12.
\end{equation}
The last result (due to the linearity of the phase function), implies
through the group compatibility condition that $\Phi_e^{\g^2}(\bi) \=
0$, and therefore, that $\g^2=\eps$, or that $\g$ is an operation of
order 2. It also implies that each of the phases $\alpha$ and $\beta$
in \rf{eq:phasesg} can be either $0$ or $1/2$, but they cannot both be
0 if $\g\neq\eps$. Thus, there can be no more than 3 operations in
$\G_e$ other than the identity.

\item Only a single element $\g\neq\eps$ in the lattice spin group
  $\G_e$ commutes with the operation $\d\in\G$, paired with $\rr$ in
  the spin point group.

\noindent {\sl\underline{Proof}:\/} If $\g\neq\eps$ in $\G_e$ commutes
with $\d$, then the relation~\rf{eq:gccGe}, between phase functions of
conjugate elements of $\g_e$, implies that
\begin{equation}
  \label{eq:commuted}
  \Phi_e^\g(\bi)\=\Phi_e^{\d\g\d^{-1}}(\rr\bi)\=\Phi_e^\g(\rr\bi).
\end{equation}
Thus, $\g$ is necessarily the operation whose phase function is given
by \rf{eq:phasesg} with $\alpha\=\beta\=1/2$.

\end{enumerate}

These lemmas, together with the facts that $G/G_\eps$ can be no bigger
than a group isomorphic to $8mm$, and that the order of the
operation $\d$ paired with $\rr$ is no bigger than 8 (proven in
section~\ref{sec:phird} below), narrow down the possible combinations
of $\G$ and $\G_e$, listed in Table~\ref{tab:extensions}, to the ones
listed in Table~\ref{tab:gammae}.

\subsection{Enumeration of spin space-group types}

We now turn to the enumeration of spin space-group types by
calculating the possible values of the phase functions for the
generators $(\rr,\d)$ and $(m,\mu)$ on the star of generating vectors
$\bi$.

\subsubsection{The phase function for \textnormal{$(\rr,\d)$}.}
\label{sec:phird}

As in the case of regular space groups for nonmagnetic crystals
\cite{rwm1,rmrw}, there is a gauge in which the phase function
$\f8(\kv)\equiv 0$ on the whole lattice. This can be shown by starting
with arbitrary values for the phase function $\f8$ and performing a
gauge transformation \rf{gauge-tr} with the gauge function
\begin{equation}
\chi(\bi)=\frac{1}{2}\Phi_{\rr}^{\d}\left(\sum_{j=i}^{i+3}{\textbf
b}^{(j)}\right),\quad i=1\ldots 4,
\end{equation}
where $\b{j}=-\b{j-4}$ for $j=5,6,7,8$.
The change to $\Phi_{\rr}^{\d}$ caused by this gauge transformation
exactly cancels it
\begin{eqnarray}\label{eq:gauge8}
\Delta\Phi_{\rr}^\d(\bi) \equiv \chi\left(\rr\bi-\bi\right) \=
\hf\Phi_{\rr}^\d(\textbf{b}^{(i+4)}-\bi)\nonumber\\
\equiv -\Phi_{\rr}^\d(\bi),
\end{eqnarray}
so that after the gauge transformation $\f8(\kv)\equiv 0$ for all wave
vectors $\kv$. Note that this implies, through the group compatibility
condition~\rf{GCC}, that $\Phi_e^{\d^8}(\kv)\=0$, so that $\d^8=\eps$,
imposing an additional restriction on the group $\G$, as
indicated in Table~\ref{tab:gammae}.

\subsubsection{The phase function for \textnormal{$(m,\mu)$}.}

When $G=8mm$ we need to calculate the additional phase function
$\fm(\kv)$, associated with the second point-group generator
$(m,\mu)$. The generating relations that contribute to the
determination of this phase function are $(m,\mu)^2=(e,\mu^2)$ and
$(\rr,\d)(m,\mu)(\rr,\d)=(m,\d\mu\d)$. Applying the group
compatibility condition \rf{GCC} to these relations, in the gauge
where $\Phi_{\rr}^\d(\kv)\=0$, yields
\begin{equation}\label{mirrorb}
\Phi_e^{\mu^2}(\bi)\=\Phi_m^\mu(m\bi + \bi),
\end{equation}
\begin{equation}\label{mirrora}
\Phi_m^{\d\mu\d}(\bi)\=\Phi_m^\mu(\rr\bi).
\end{equation}

We shall first determine the value of the phase $\Phi_m^\mu(\b1)$
using Eq.~\(mirrorb), and then use Eq.~\(mirrora) to infer the values
of $\Phi_m^\mu$ on the remaining three generating vectors. We start by
noting that $\mu^2$ is an operation in $\G_e$ which is the square of
an operation in $\G$. Inspection of all the possibilities, listed in
Table~\ref{tab:gammae}, reveals that only two operations, $2_\zb$ and
$\eps$, satisfy this condition. Furthermore, if $m$ is the mirror that
leaves $\b1$ invariant, then application of Eq.~\rf{mirrorb} to
${\textbf b}^{(3)}$ which is perpendicular to $m$ (\mbox{$m\b3=-\b3$})
yields
\begin{equation}\label{mirrorb3}
\Phi_e^{\mu^2}(\b3)\=\Phi_m^\mu(m\b3+\b3)\=0.
\end{equation}
This implies that $\mu^2$ cannot be $\sr2$, because $\Phi_e^{\sr2}$
has the value $\hf$ on all lattice generating vectors. Therefore,
$\mu^2$ must be equal to $\eps$. Application of Eq.~\rf{mirrorb} to
${\textbf b}^{(1)}$ now yields
\begin{equation}\label{mirrorb1}
0\=2\Phi_m^\mu(\b1) \Longrightarrow
  \Phi_m^\mu(\b1) \= 0 {\rm \ or\ } \frac12,
\end{equation}
and application of Eq.~\rf{mirrorb} to ${\textbf b}^{(2)}$ and
${\textbf b}^{(4)}$ shows that $\Phi_m^\mu(\b2)\=\Phi_m^\mu(\b4)$, but
provides no further information regarding the actual values of these
phases.

Next, we examine Eq.~\rf{mirrora}, which can be rephrased (using the
group compatibility condition \(GCC)) as
\begin{equation}
\Phi_m^\mu(\bi)+\Phi_e^{\mu^{-1}\d\mu\d}(\bi)\=\Phi_m^\mu(\rr\bi).
\end{equation}
The value of $\Phi_m^\mu$ on $\b1$ determines the values of
$\Phi_m^\mu$ on the remaining generating vectors through some phase
function, associated with an element of $\G_e$. Note that
$\mu^{-1}\d\mu\d$ is an operation in $\G_e$ which is the product of
two operations, $\mu^{-1}\d\mu$ and $\d$, that are conjugate in $\G$.
Inspection of Table~\ref{tab:gammae} shows that if the product of any
two conjugate operation in $\G$ is in $\G_e$, then this product is
necessarily either $2_\zb$ or the identity $\eps$. Substituting the
values $\Phi_e^{\eps}(\bi)\equiv0000$ and
$\Phi_e^{2_\zb}(\bi)\equiv\hf\hf\hf\hf$, we conclude that
\begin{equation}\label{mirrorfinal}
\Phi_m^\mu(\b{i}) \= 
\begin{cases}
  0000 \text{\ or\ } \hf\hf\hf\hf & \text{if $\d\mu\d=\mu$},\\
  0\hf0\hf \text{\ or\ } \hf0\hf0 & \text{if $\d\mu\d=\mu 2_\zb$}.
\end{cases}
\end{equation}
Thus, there are two spin space groups for each 2-dimensional octagonal
spin point group with $G=8mm$.

\subsection{Spin group tables}
\label{sec:notation}

The resulting 2-dimensional octagonal spin point groups and spin space
groups are listed in Table \ref{2d8groups} for $G=8$, and in Table
\ref{2d8mmgroups} for $G=8mm$, using the following notation:

Each line in the Tables represents one or more spin point groups and
their associated spin space groups. The spin point groups are given by
their generators, listed in the fifth column of each Table. The first
four columns provide the group theoretic structure of the spin point
group by listing the lattice spin group $\G_e$, the normal subgroup
$G_\eps$ of $G$, paired with $\G_e$, the quotient group $G/G_\eps$,
and the full group of spin-space rotations $\G$, satisfying the
requirement that $G/G_\eps\simeq\G/\G_e$. We use stars and daggers to
denote optional primes on elements of $\G$ (\ie\ the application of
time inversion after a spin space rotation). If two operations in $\G$
can be independently primed or unprimed we use a star for the first
and a dagger for the second. For example, the symbol $2^*2^*2$
stands for the two possible groups $\G=222$ and $2'2'2$, whereas the
symbol $\ax4$ stands for four distinct groups, $\G=422$, $4'22'$,
$42'2'$, and $4'2'2$.

To list the spin space groups for each spin point group we must
specify the values of the phase functions for all the spin point-group
generators on the four generating vectors of the lattice. The phase
functions $\feg$ for generators of the form $(e,\g)$ are already
listed in Table~\ref{tab:gammae} and are not repeated in Tables
\ref{2d8groups} and \ref{2d8mmgroups}. The phase function $\f8$ is
zero everywhere due to the choice of gauge, and is therefore also not
listed in Tables \ref{2d8groups} and \ref{2d8mmgroups}. The two
possible values of the phase function $\fm$ when the point group is
$8mm$, which according to Eq.~\rf{mirrorfinal} depend on the value of
$\d\mu\d$, are listed in the sixth column of Table \ref{2d8mmgroups}.
If $\d\mu\d=\mu$ we write ``$0;\hf$'' to indicate that
$\fm(\bi)\=0000$ or $\hf\hf\hf\hf$. When $\d\mu\d\neq\mu$ we write
``$A$'' to indicate that $\fm(\bi)\=0\hf 0\hf$ or $\hf 0\hf 0$,
alternating its value from one generating vector to the next.

In the last column of each Table we give a unique symbol for each spin
space group, based on the familiar International (Hermann-Mauguin)
symbols for the regular (nonmagnetic) space groups. To incorporate all
the spin space-group information we augment the regular symbol in the
following ways: (1) The symbol for the lattice spin group $\G_e$ is
added in parentheses immediately after the regular space group symbol,
unless $\G_e=1$.  (2) In the case of 2-dimensional octagonal spin
space groups, the values of the phase functions associated with the
elements of $\G_e$ are unique and therefore need not be listed. In
general, one can encode these phase functions by indicating the
sublattice $L_0$ (for which $L/L_0$ is isomorphic to $\G_e$, as
described in section~\ref{sec:iso}) as a subscript of the magnetic
lattice symbol $P$. (3) To each generator of the point group $G$ we
add a superscript listing an operation from the coset of $\G_e$ with
which it is paired (if that operation is $\eps$ we omit it, if it is
$\eps'$ we simply add a prime, we use stars and daggers, as described
above, to denote multiple groups, and we omit the axis about which
rotations are performed if it is the $\zb$-axis).  (4) The value of
phase function $\fm$, when the point group is $8mm$, is encoded by
replacing the secondary $m$ by a $b$ (as in the International symbols)
when $\fm(\bi)\equiv\hf\hf\hf\hf$, and by adding a subscript $a$ (for
``alternating'') so that $m_a$ indicates that $\fm(\bi)\equiv0\hf0\hf$
and $b_a$ indicates that $\fm(\bi)\equiv\hf0\hf0$.

\subsection{Selection rules due to \textnormal{$\G_e$}}
\label{sec:sr}

Operations in $G$ which impose selection rules for neutron diffraction
are those that leave some lattice vectors invariant.  In two
dimensions these can be the identity $e$, which leaves all lattice
vectors invariant or mirror lines that leave all vectors along them
invariant. We first consider the selection rules that arise from
operations $(e,\g)$ where $\g\in\G_e$, and therefore apply to all
lattice vectors, expressed in terms of the four generating vectors as
$\kv=n_1\b1 + n_2\b2 + n_3\b3 + n_4\b4$.

\subsubsection{Selection rules for \textnormal{$\G_e=2,2',1'$}.}

Denoting the generator of $\G_e$ by $\g$, the phases
$\Phi_e^\g(\bi)\=\hf\hf\hf\hf$ in all of these cases. This implies
through the eigenvalue relation \rf{eigen} that the form of $\sv(\kv)$
depends on $\g$ and on the parity of $\sum n_i$ as follows
\begin{equation}
  \label{eq:sr1}
  \g\sv(\kv) = e^{-i\pi \sum n_i}\sv(\kv).
\end{equation}
Namely, whenever $\sum n_i$ is even the phase in Eq.~\rf{eq:sr1}
vanishes and $\sv(\kv)$ must be invariant under the operation $\g$;
and whenever $\sum n_i$ is odd the phase is $i\pi$ and $\sv(\kv)$ must
change its sign under $\g$. The consequences for the three possible
operations $\g$ are summarized in Table \ref{tab:sr1}.

\subsubsection{Selection rules for \textnormal{$\G_e=222,22'2'$}.}

Here $\G_e$ is generated by $(e,2^*_\xb)$ and ($e,2^*_\yb$), with
phase functions given by $\Phi_e^{2^*_\xb}(\b{i})\=0\hf0\hf$ and
$\Phi_e^{2^*_\yb}(\b{i})\=\hf0\hf0$. The eigenvalue relations
\rf{eigen} for the two generators are
\begin{eqnarray}
  \label{eq:sr2a}
  2^*_\xb\sv(\kv) &= &e^{-i\pi (n_2 + n_4)}\sv(\kv),\\
  \label{eq:sr2b}
  2^*_\yb\sv(\kv) &= &e^{-i\pi (n_1 + n_3)}\sv(\kv),
\end{eqnarray}
so that $\sv(\kv)$ remains invariant (changes its sign) under
$2^*_\xb$ if $n_2 + n_4$ is even (odd); and remains invariant (changes
its sign) under $2^*_\yb$ if $n_1 + n_3$ is even (odd). These results
are summarized in Table~\ref{tab:sr2} for the two possible $\G_e$'s.

\subsection{Selection rules on mirror lines}

In addition to the selection rules arising from $\G_e$ there are also
selection rules that occur when $\kv$ lies on one of the mirror lines
and is therefore invariant under reflection through that particular
mirror. In this case the eigenvalue equation \rf{eigen} imposes
further restrictions on the Fourier coefficients of the spin density
field.

\begin{figure}
\caption{Generating vectors and mirror lines for
  two-dimensional octagonal lattices. The solid arrows are the star of
  generating vectors and their negatives $\pm\b1\ldots\pm\b4$. The
  dashed lines show the two types of mirrors in the $8mm$ point group,
  as described in the text.}
\label{fig:star}
\resizebox{!}{7cm}{\includegraphics{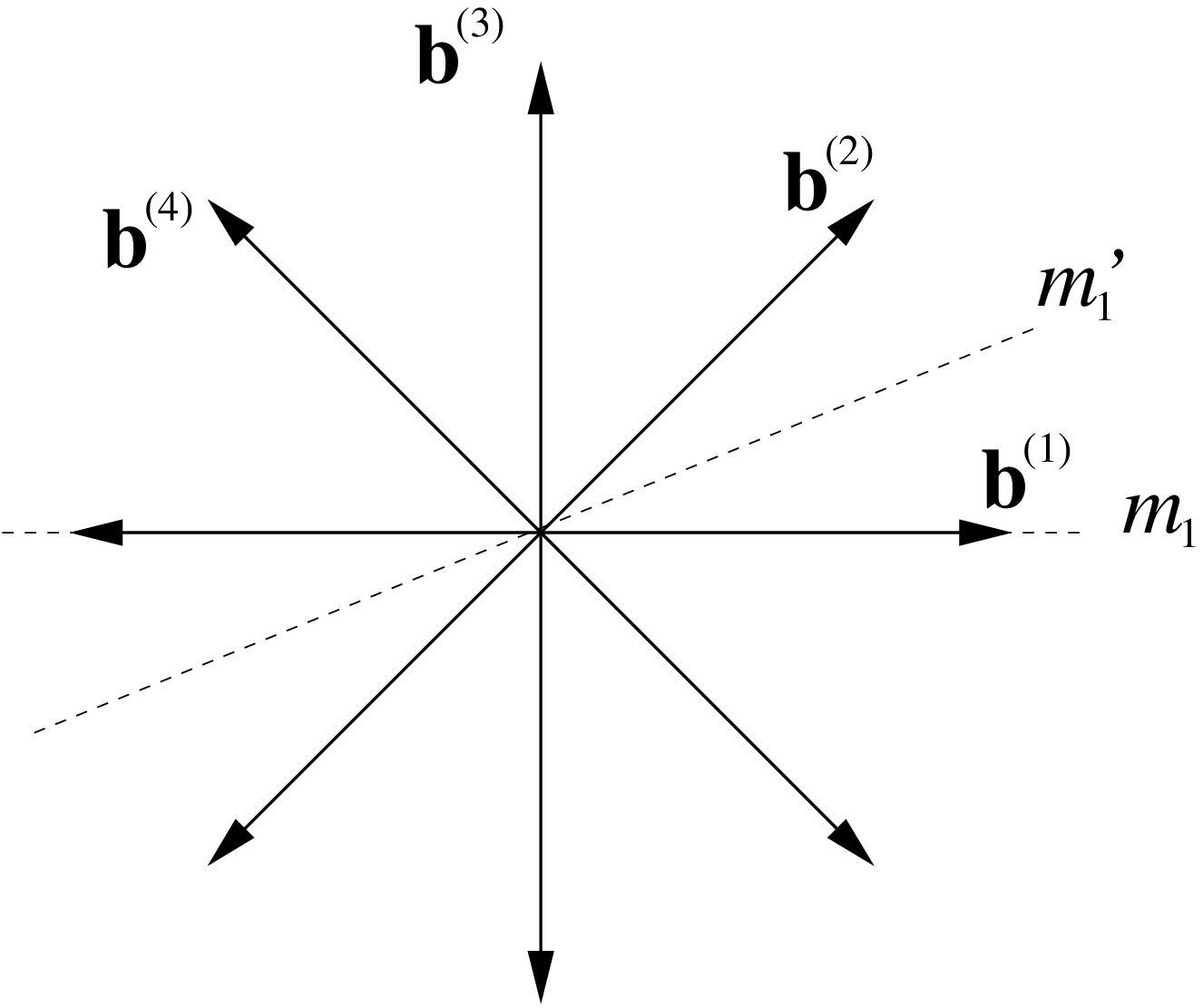}}
\end{figure}

Vectors lying along the mirror $m_i$, that leaves the generating
vector $\b{i}$ invariant, have the general form
\begin{equation}
  \label{eq:kalong}
  \kv_i=n_i\bi+l_i(\b{i-1}+\b{i+1}),\quad i=1,2,3,4,  
\end{equation}
where all indices are taken modulo 8, and $\b{j}=-\b{j-4}$ for
$j=5,6,7,8$.  Selection rules along $m_1$, which is the mirror $m$
used to generate the point group (see Figure~\ref{fig:star}), are
determined by the equation
\begin{equation} \label{eq:mk1}
\mu\sv(\kv_1)=e^{-2i\pi n_1\fm(\b1)}\sv(\kv_1),
\end{equation}
where we have used the fact [Eq.~\rf{mirrorfinal}] that
$\fm(\b2)-\fm(\b4)\=0$. Therefore, the form of $\sv(\kv_1)$ depends on
$\mu$, on the parity of $n_1$, and on the phase $\fm(\b1)$ as follows:
If $n_1$ is odd {\it and\/} $\fm(\b1)\equiv\hf$ then $\sv(\kv_1)$ must
change its sign under $\mu$, otherwise $\sv(\kv_1)$ must remain
invariant under $\mu$.

To obtain the selection rules for vectors lying along the remaining
mirrors $m_i$ $(i=2,3,4)$, we use successive applications of the
symmetry operation $(r,\d)$ to the result~\rf{eq:mk1} for $m_1$.
Since $m\kv_1=\kv_1$, it follows from relation~\rf{eq:conj}, between
phase functions of conjugate operations, that
\begin{equation}
\d^{i-1}\mu\d^{1-i}\sv(\kv_i)=e^{-2i\pi n_i\fm(\b1)}\sv(\kv_i).
\end{equation}
Thus, in general, for a vector $\kv_i$, given by Eq.~\rf{eq:kalong}
and lying along the mirror $m_i$, the form of $\sv(\kv_i)$ must
satisfy the following requirement: If $n_i$ is odd {\it and\/}
$\fm(\b1)\equiv\hf$ then $\sv(\kv_i)$ must change its sign under
$\d^{i-1}\mu\d^{1-i}$, otherwise $\sv(\kv_i)$ must remain invariant
under $\d^{i-1}\mu\d^{1-i}$. Note that $\fm(\b1)\equiv0$ for spin
space groups of type $P8^\d m^\mu m(\G_e)$ and $P8^\d m_a^\mu
m(\G_e)$, and that $\fm(\b1)\equiv\hf$ for spin space groups of type
$P8^\d b^\mu m(\G_e)$ and $P8^\d b_a^\mu m(\G_e)$. Also note that in
most cases $\G$ is abelian, so $\d^{i-1}\mu\d^{1-i}=\mu$, and the
selection rules take a much simpler form. These results are summarized
in the second column of Table~\ref{tab:srmirrors}.

Vectors lying along the mirror $m'_i$, which is between the generating
vector $\b{i}$ and $\b{i+1}$, have the general form
\begin{equation}
  \label{eq:kbetween}
  \kv'_i=n_i(\bi+\b{i+1})+l_i(\b{i-1}+\b{i+2}),\quad i=1,2,3,4,
\end{equation}
again with all indices taken modulo 8, and $\b{j}=-\b{j-4}$ for
$j=5,6,7,8$.  Using the group compatibility condition for the relation
$m'_1=\rr m_1$ (see Figure~\ref{fig:star}), in the gauge where
$\f8(\kv)\equiv 0$, yields
\begin{equation}
\Phi_{m'_1}^{\d\mu}(\kv'_1)\equiv
\fm(\kv'_1)\equiv (n_1+l_1)\left(\fm(\b1)+\fm(\b2)\right),
\end{equation} 
where we have used the fact [Eq.~\rf{mirrorfinal}] that
$\fm(\b3)\=\fm(\b1)$ and $\fm(-\b4)\=\fm(\b2)$. Therefore, the
selection rules for $m'_1$ are determined
by the equation
\begin{equation}
\d\mu\sv(\kv'_1) = 
e^{-2i\pi (n_1+l_1)\left(\fm(\b1)+\fm(\b2)\right)}\sv(\kv'_1),
\end{equation} 
and again, by successive rotations $(r,\d)$, we obtain the
selection rules for the remaining mirrors $m'_i$, 
\begin{equation}
\d^i\mu\d^{1-i}\sv(\kv'_i) = 
e^{-2i\pi (n_i+l_i)\left(\fm(\b1)+\fm(\b2)\right)}\sv(\kv'_i).
\end{equation} 
Thus, in general, for a vector $\kv'_i$, given by Eq.~\rf{eq:kbetween}
and lying along the mirror $m'_i$, the form of $\sv(\kv'_i)$ must
satisfy the following requirement: If $n_i+l_i$ is odd {\it and\/}
$\fm(\b1)+\fm(\b2)\equiv\hf$ then $\sv(\kv'_i)$ must change its sign
under $\d^i\mu\d^{1-i}$, otherwise $\sv(\kv'_i)$ must remain invariant
under $\d^i\mu\d^{1-i}$. Note that $\fm(\b1)+\fm(\b2)\equiv0$ for spin
space groups of type $P8^\d m^\mu m(\G_e)$ and $P8^\d b^\mu m(\G_e)$,
and that $\fm(\b1)+\fm(\b2)\equiv\hf$ for spin space groups of type
$P8^\d m_a^\mu m(\G_e)$ and $P8^\d b_a^\mu m(\G_e)$. Also note
that in most cases $\G$ is abelian, so $\d^i\mu\d^{1-i}=\d\mu$,
and the selection rules take a much simpler form. These results are
summarized in the third column of Table~\ref{tab:srmirrors}. 

\ack{The authors thank Benji Fisher for his comments on a draft of
  this manuscript. This research was funded by the Israel Science
  Foundation through Grant No.~278/00.}

\referencelist[spin]
%\clearpage

%-------------- Table 1 -------------------------------------------
\onecolumn
\begin{table}
\caption{\label{tab:extensions}
    Possible lattice spin groups $\G_e$ and their extensions into the full
    groups $\G$ of spin-space operations. All possible $\G_e$'s are listed
    in the first column. The second column shows the constraints on $\G$
    imposed by the fact that $\G_e$ is a normal subgroup of $\G$. The
    integer $k$ is arbitrary.} 
  \setlength{\extrarowheight}{2pt}
  \begin{center}
    \begin{tabular}{>{$}c<{$}|>{$}c<{$}}
      \cline{1-2}
      \G_e&\G\\ 
      \cline{1-2}
      1;1'&\G\subseteq SO(3)\cross 1'\\
      n;n';n1'&\G\subseteq (kn)221'\\
      222;2'2'2'&\G\subseteq 4321'\\
      2_\zb2'2'&\G\subseteq 4221'\\
    \end{tabular}
  \end{center}
\end{table}

%-------------- Table 2 --------------------------------------------

\begin{table}
\caption{Normal subgroups $G_\eps$ of the point groups $G=8mm$ and
    $8$. The resulting quotient group $G/G_\eps$ is represented in
    the third column by a point group, isomorphic to it. Constraints on
    the spin-space orperations $\d$ and $\mu$, paired with the
    generators $\rr$ and $m$ of $G$ are listed in the fourth
    column. In each line the first power of $\d$ that is in $\G_e$ is
    given. $\mu^2$ is always in $\G_e$, therefore we only note whether
    $\mu\in\G_e$. If $\d$ or $\mu$ are in $\G_e$ they are chosen as
    $\eps$.} 
\label{tab:normal}
\setlength{\extrarowheight}{2pt}
\begin{center}
\begin{tabular}{>{$}c<{$}>{$}c<{$}>{$}c<{$}|>{$}c<{$}}
\cline{1-4}
G &G_\eps &G/G_\eps &\textnormal{Constraints}\\
\cline{1-4}
8mm&8mm&1&\d=\eps,\mu=\eps\\
&8&m&\d=\eps,\mu\notin\G_e\\
&4mm&2&\d^2\in\G_e,\mu=\eps\\
&4m'm'&2&\d=\mu\notin\G_e\\
&4&2mm&\d^2\in\G_e,\mu\notin\G_e,\d\G_e\neq\mu\G_e\\
&2&4mm&\d^4\in\G_e,\mu\notin\G_e,\d^2\G_e\neq\mu\G_e\\
&1&8mm&\d^8\in\G_e,\mu\notin\G_e,\d^4\G_e\neq\mu\G_e\\
\cline{1-4}
8  &8 &1 &\d=\eps\\
&4 &2 &\d^2\in\G_e\\
&2 &4 &\d^4\in\G_e\\
&1 &8 &\d^8\in\G_e\\
\end{tabular}
\end{center}
\end{table}

%-------------- Table 3 -------------------------------------------

\begin{table}
  \caption{
    Possible lattice spin groups $\G_e$ and their extensions
    into the full 
    groups $\G$ of spin-space operations, compatible with
    the 2-dimensional rank-4 octagonal lattice. All possible $\G_e$'s
    are listed in the first column and constraints on the possible
    supergroups $\G$ are listed in the second column. The phase
    functions for the generators of $\G_e$ are listed in the third column. }
  \label{tab:gammae}
  \setlength{\extrarowheight}{2pt}
  \begin{center}
    \begin{tabular}{>{$}c<{$}|>{$}l<{$}|>{$}l<{$}}
      \cline{1-3}
      \G_e & {\rm Constraints\ on\ } \G & {\rm Phase\ functions\ for\
      generators\ of\ } \G_e\\
      \cline{1-3}
      1   & \G_e\subseteq\G\subset 8221'  & N/A \\
      1'  & \G_e\subseteq\G\subseteq 8221'
        & \Phi_e^{\eps'}(\bi)\equiv\hf\hf\hf\hf\\
      2   & \G_e\subseteq\G\subset 8221'
        & \Phi_e^{2_\zb}(\bi)\equiv\hf\hf\hf\hf\\
      2'  & \G_e\subseteq\G\subseteq 8221'
        & \Phi_e^{2_\zb'}(\bi)\equiv\hf\hf\hf\hf\\
      222 & 422\subseteq\G\subseteq 4221'
        & \Phi_e^{2_\xb}(\bi)\equiv0\hf0\hf;\
          \Phi_e^{2_\yb}(\bi)\equiv\hf0\hf0\\
      2_\zb2'2' & 42'2'\subseteq\G\subseteq 4221'
        & \Phi_e^{2_\xb'}(\bi)\equiv0\hf0\hf;\
          \Phi_e^{2_\yb'}(\bi)\equiv\hf0\hf0\\
    \end{tabular}
  \end{center}
\end{table}

%-------------- Table 4 -------------------------------------------

\begin{table}
\caption{Two-dimensional octagonal spin point groups and spin space
  group types with point group $G=8$. The phase functions $\f8$ is zero
  everywhere by choice of gauge. The values of the phase functions
  $\Phi_e^{\g}$ for $\g\in\G_e$ on the lattice generating vectors are
  listed in Table~\ref{tab:gammae}. The symbols for the spin
  space groups are listed in the rightmost column using the notation
  described in section~\ref{sec:notation}.}
\label{2d8groups}
\setlength{\extrarowheight}{2pt}
\begin{center}
\begin{tabular}{>{$}c<{$}>{$}c<{$}>{$}c<{$}>{$}l<{$}|>{$}l<{$}|>{$}l<{$}}
%\hline
\cline{1-6}
\G_e&G_\eps&G/G_\eps&\G&\textnormal{Generators}&\textnormal{Space Groups}\\
\cline{1-6}
 1&8&1&1&(\rr,\eps)&P8\\
%\cline{2-5}
 &4&2&2^*&(\rr,2^*_\zb)&P8^{2^*}\\
%\cline{4-5}
 & & &1'&(\rr,\eps')&P8'\\
%\cline{2-5}
 &2&4&4^*&(\rr,4_\zb^{*})&P8^{4^{*}}\\
%\cline{2-5}
 &1&8&8^*&(\rr,8_\zb^{*})&P8^{8^{*}}\\
 & & & &(\rr,8_\zb^{3*})&P8^{8^{3*}}\\
%\hline
\cline{1-6}
 2&8&1&2&(\rr,\eps),(e,2_\zb)&P8(2)\\
%\cline{2-5}
 &4&2&4^*&(\rr,4_\zb^*),(e,2_\zb)&P8^{4^*}(2)\\
%\cline{4-5}
 & & &2^*2^*2&(\rr,2_\xb^*)(e,2_\zb)&P8^{2^*_\xb}(2)\\
%\cline{4-5}
 & & &21'&(\rr,\eps')(e,2_\zb)&P8'(2)\\
%\cline{2-5}
 &2&4&8^*&(\rr,8_\zb^{*})(e,2_\zb)&P8^{8^{*}}(2)\\
%\hline
\cline{1-6}
 2'&8&1&2'&(\rr,\eps)(e,2_\zb')&P8(2')\\
%\cline{2-5}
 &4&2&2'2'2&(\rr,2^*_\xb)(e,2'_\zb)&P8^{2^*_\xb}(2')\\
%\cline{4-5}
 & & &21'&(\rr,\eps')(e,2'_\zb)&P8'(2')\\
%\cline{2-5}
 &2&4&41'&(\rr,\sr4)(e,2'_\zb)&P8^{4}(2')\\
%\cline{2-5}
 &1&8&81'&(\rr,\sr8)(e,2'_\zb)&P8^{8}(2')\\
%\hline
\cline{1-6}
 1'&8&1&1'&(\rr,\eps)(e,\eps')&P8(1')\\
%\cline{2-5}
 &4&2&21'&(\rr,2_\zb)(e,\eps')&P8^2(1')\\
%\cline{2-5}
 &2&4&41'&(\rr,4_\zb)(e,\eps')&P8^{4}(1')\\
%\cline{2-5}
 &1&8&81'&(\rr,8_\zb)(e,\eps')&P8^{8}(1')\\
& & & &(\rr,8_\zb^{3})(m,\eps')&P8^{8^{3}}(1')\\
\cline{1-6}
%\hline
222&4&2&4^*22^*&(\rr,4_\zb^*)(e,2_\xb)(e,2_\yb)&P8^{4^*}(222)\\
\cline{1-6}
\end{tabular}
\end{center}
\end{table}

%-------------- Table 5 -------------------------------------------

\clearpage
\onecolumn
\setlength{\extrarowheight}{2pt}
\begin{longtable}{|>{$}c<{$}|>{$}l<{$}|>{$}l<{$}|>{$}l<{$}|>{$}l<{$}|>{$}c<{$}|>{$}l<{$}>{$}l<{$}|}
\caption{Two-dimensional octagonal spin point groups and spin space
  group types with point group $G=8mm$. The phase functions $\f8$ is zero
  everywhere by choice of gauge. The values of the phase functions
  $\Phi_e^{\g}$ for $\g\in\G_e$ on the lattice generating vectors are
  listed in Table~\ref{tab:gammae}. The possible values of the phase
  function $\fm$ are listed in the sixth column using the notation
  described in section~\ref{sec:notation}. The symbols for the spin
  space groups are listed in the rightmost column using the notation
  which is also described in section~\ref{sec:notation}.}
\label{2d8mmgroups}\\
\hline
\G_e&G_\eps&G/G_\eps&\G&\textnormal{Generators}&\Phi_m^\mu&\multicolumn{2}{l|}{Spin
  Space Group Types}\\ 
\hline
\endfirsthead\\
\caption{\textnormal{continued}}\\
\hline
\G_e&G_\eps&G/G_\eps&\G&\textnormal{Generators}&\Phi_m^\mu&\multicolumn{2}{l|}{Spin
  Space Group Types}\\
\hline 
\endhead 
\hline 
\multicolumn{8}{r}{\textnormal{continued on next page}} 
\endfoot
\hline
\endlastfoot
\hline 1&8mm&1&1&(\rr,\eps)(m,\eps)&0;\hf&P8mm;&P8bm\\
\cline{2-8} 
&8&2&2^*&(\rr,\eps)(m,2_\zb ^*)&0;\hf&P8m^{2^*}m;&P8b^{2^*}m\\
\cline{4-8} 
& & &1'&(\rr,\eps)(m,\eps')&0;\hf&P8m'm;&P8b'm\\
\cline{2-8} 
&4mm&2&2^*&(\rr,2_\zb^*)(m,\eps)&0;\hf&P8^{2^*}mm;&P8^{2^*}bm\\
\cline{4-8} 
& & &1'&(\rr,\eps')(m,\eps)&0;\hf&P8'mm;&P8'bm\\
\cline{2-8} 
&4m'm'&2&2^*&(\rr,2_\zb^*)(m,2_\zb^*)&0;\hf&P8^{2^*}m^{2^*}m;&P8^{2^*}b^{2^*}m\\
\cline{4-8} 
& & &1'&(\rr,\eps')(m,\eps')&0;\hf&P8'm'm;&P8'b'm\\
\cline{2-8} 
&4&2mm&2^*2^\dagger 2^{*\dagger}&(\rr,2_\zb^*)(m,2_\xb^\dagger)&0;\hf&
P8^{2^*}m^{2^\dagger_\xb}m; &P8^{2^*}b^{2^\dagger_\xb}m\\
\cline{4-8} 
& & &21'&(\rr,2_\zb^{*})(m,2{'}_\zb^{*})&0;\hf 
&P8^{2^*}m^{2'^*}m; &P8^{2^*}b^{2'^*}m\\
\cline{5-8}
 & & & &(\rr,2_\zb^*)(m,\eps')&0;\hf &P8^{2^*}m'm; &P8^{2^*}b'm\\
\cline{5-8} 
& & & &(\rr,\eps')(m,2_\zb^*)&0;\hf &P8'm^{2^*}m; &P8'b^{2^*}m\\
\cline{2-8} 
&2&4mm&4^*2^\dagger 2^{*\dagger}&(\rr,4^{*}_\zb)(m,2_\xb^\dagger)&0;\hf
&P8^{4^{*}}m^{2^\dagger}m; &P8^{4^{*}}b^{2^\dagger}m\\
\cline{2-8} 
&1&8mm&8^*2^\dagger2^{*\dagger}&(\rr,8_\zb^{*})(m,2_\xb^\dagger)&0;\hf
&P8^{8^{*}}m^{2^\dagger_\xb}2; &P8^{8^{*}}b^{2^\dagger_\xb}m\\
\cline{5-8} 
& & & &(\rr,8_\zb^{3*})(m,2_\xb^\dagger)&0;\hf
&P8^{8^{3*}}m^{2^\dagger_\xb}2; &P8^{8^{3*}}b^{2^\dagger_\xb}m\\

\hline 2&8mm&1&2&(\rr,\eps),(m,\eps)(e,2_\zb)&0;\hf
&P8mm(2); &P8bm(2)\\
\cline{2-8} &8&m&2^*2^*2&(\rr,\eps)(m,2_\xb^*)(e,2_\zb)&0;\hf
&P8m^{2_\xb^*}m(2); &P8b^{2_\xb^*}m(2)\\
\cline{4-8} & & &21'&(\rr,\eps)(m,\eps')(e,2_\zb)&0;\hf
&P8m'm(2); &P8b'm(2)\\
\cline{2-8} &4mm&2&2^*2^*2&(\rr,2_\xb^*)(m,\eps)(e,2_\zb)&0;\hf
&P8^{2_\xb^*}mm(2); &P8^{2_\xb^*}bm(2)\\
\cline{4-8} &&&21'&(\rr,\eps')(m,\eps)(e,2_\zb)&0;\hf
&P8'mm(2); &P8'bm(2)\\
\cline{4-8} &&&4^*&(\rr,4^*_\zb)(m,\eps)(e,2_\zb)&A
&P8^{4^*}m_am(2); &P8^{4^*}b_am(2)\\
\cline{2-8} &4m'm'&2&2^*2^*2&(\rr,2_\xb^*)(m,2_\xb^*)(e,2_\zb)&0;\hf
&P8^{2^*_\xb}m^{2^*_\xb}m(2); &P8^{2^*_\xb}b^{2^*_\xb}m(2)\\
\cline{4-8} &&&21'&(\rr,\eps')(m,\eps')(e,2_\zb)&0;\hf
&P8'm'm(2); &P8'b'm(2)\\
\cline{2-8} &4&2mm&4^*2^\dagger 2^{*\dagger}&(\rr,4^*_\zb)(m,2^\dagger_\xb)(e,2_\zb)&0;\hf&
P8^{4^*}m^{2^\dagger_\xb}m(2); &P8^{4^*}b^{2^\dagger_\xb}m(2)\\
\cline{5-8} &&&&(\rr,2_\xb^\dagger)(m,2^*_\xyb)(e,2_\zb)&A&
P8^{2_\xb^\dagger}m_a^{2^*_\xyb}m(2); &P8^{2_\xb^\dagger}b_a^{2^*_\xyb}m(2)\\
\cline{4-8} &&&2221'&(\rr,\eps')(m,2_\xb^*)(e,2_\zb)&0;\hf&
P8'm^{2^*_\xb}m(2); &P8'b^{2^*_\xb}m(2)\\
\cline{5-8}&&&&(\rr,2_\xb^*)(m,\eps')(e,2_\zb)&0;\hf
&P8^{2^*_\xb}m'm(2); &P8^{2^*_\xb}b'm(2)\\
\cline{5-8}&&&&(\rr,2_\xb^{*})(m,2_{\xb}^{*'})(e,2_\zb)&0;\hf&
P8^{2^*_\xb}m^{2'^*_\xb}m(2); &P8^{2^*_\xb}b^{2'^*_\xb}m(2)\\
\cline{2-8} &2&4mm&8^*2^\dagger 2^{*\dagger}
&(\rr,8_\zb^{*})(m,2_\xb^\dagger)(e,2_\zb)&0;\hf
&P8^{8^{*}}m^{2^\dagger_\xb}m(2); &P8^{8^{*}}b^{2^\dagger_\xb}m(2)\\

\pagebreak

\hline 2'&8mm&1&2'&(\rr,\eps)(m,\eps)(e,2_\zb')&0;\hf
&P8mm(2'); &P8bm(2')\\
\cline{2-8} &8&m&2'22'&(\rr,\eps)(m,2_\xb)(e,2_\zb')&0;\hf
&P8m^{2_\xb}m(2'); &P8b^{2_\xb}m(2')\\
\cline{4-8} &&&21'&(\rr,\eps)(m,\eps')(e,2_\zb')&0;\hf&
P8m'm(2'); &P8b'm(2')\\
\cline{2-8} &4mm&2&2'22'&(\rr,2_\xb)(m,\eps)(e,2_\zb')&0;\hf
&P8^{2_\xb}mm(2'); &P8^{2_\xb}bm(2')\\
\cline{4-8} &&&21'&(\rr,\eps')(m,\eps)(e,2_\zb')&0;\hf
&P8'mm(2'); &P8'bm(2')\\
\cline{2-8} &4m'm'&2&2'22'&(\rr,2_\xb^*)(m,2_\xb)(e,2_\zb')&0;\hf&
P8m^{2^*_\xb}m^{2^*_\xb}m(2'); &P8m^{2^*_\xb}b^{2^*_\xb}m(2')\\
\cline{4-8} &&&21'&(\rr,\eps')(m,\eps')(e,2_\zb')&0;\hf
&P8'm'm(2'); &P8'b'm(2')\\
\cline{2-8} &4&2mm&2221'&(\rr,\eps')(m,2_\xb^*)(e,2_\zb')&0;\hf
&P8'm^{2_\xb}m(2'); &P8'b^{2_\xb}m(2')\\
\cline{5-8}&&&&(\rr,2_\xb^*)(m,\eps')(e,2_\zb')&0;\hf&
P8^{2_\xb}m'm(2'); &P8^{2_\xb}b'm(2')\\
\cline{2-8} & 2&4mm&4221'&(\rr,4_\zb^*)(m,2_\xb)(e,2_\zb')&0;\hf
&P8^{4^*}m^{2_\xb}m(2'); &P8^{4^*}m^{2_\xb}m(2')\\
\cline{2-8}
&1&8mm&8221'&(\rr,8_\zb^{*})(m,2_\xb)(e,2'_\zb)&0;\hf&
P8^{8^{*}}m^{2_\xb}m(2'); &P8^{8^{*}}b^{2_\xb}m(2')\\

\hline 1'&8mm&1&1'&(\rr,\eps)(m,\eps)(e,\eps')&0;\hf&
P8mm(1'); &P8bm(1')\\
\cline{2-8} &8&m&21'&(\rr,\eps)(m,2_\zb)(e,\eps')&0;\hf&
P8m^2m(1'); &P8b^2m(1')\\
\cline{2-8} &4mm&2&21'&(\rr,2_\zb)(m,\eps)(e,\eps')&0;\hf
&P8^2mm(1'); &P8^2bm(1')\\
\cline{2-8} &4m'm'&2&21'&(\rr,2_\zb)(m,2_\zb)(e,\eps')&0;\hf&
P8^2m^2m(1'); &P8^2b^2m(1')\\
\cline{2-8} &4&2mm&2221'&(\rr,2_\zb)(m,2_\xb)(e,\eps')&0;\hf
&P8^2m^{2_\xb}m(1'); &P8^2b^{2_\xb}m(1')\\
\cline{2-8} &2&4mm&4221'&(\rr,4_\zb)(m,2_\xb)(e,\eps')&0;\hf
&P8^{4}m^{2_\xb}m(1'); &P8^{4}b^{2_\xb}m(1')\\
\cline{2-8} &1&8mm&8221'&(\rr,8_\zb)(m,2_\xb)(e,\eps')&0;\hf
&P8^{8}m^{2_\xb}m(1'); &P8^{8}b^{2_\xb}m(1')\\
&&&&(\rr,8_\zb^{3})(m,2_\xb)(e,\eps')&0;\hf&
P8^{8^{3}}m^{2_\xb}m(1'); &P8^{8^{3}}b^{2_\xb}m(1')\\
\hline
222&4mm&2&4^*22^*&(\rr,4_\zb^*)(m,\eps)(e,2_\xb)(e,2_\yb)&A
&P8^{4^*}m_am(222); &P8^{4^*}b_am(222)\\
\cline{2-8} &4&2mm&4221'&(\rr,4_\zb)(m,\eps')(e,2_\xb)(e,2_\yb)&A
&P8^{4}m'_am(222); &P8^{4}b'_am(222)\\
\hline
\end{longtable}
\clearpage
%\twocolumn

%-------------- Table 6 -------------------------------------------

\begin{table}
\caption{Restrictions on the form of $\sv(\kv)$ for any wave vector
  $\kv$ in the magnetic lattice when $\G_e=2$, $2'$, or $1'$. In each
  case the form of $\sv(\kv)$ depends on the parity of $\sum n_i$
  where $\kv=\sum n_i\bi$. Colors refer to the points in
  Figure~\ref{fig:srplot}. }  
\label{tab:sr1}
\setlength{\extrarowheight}{3pt}
\begin{center}
\begin{tabular}{>{$}c<{$}|>{$}c<{$}>{$}c<{$}}
\cline{1-3}
&\sum n_i\textnormal{ even }&\sum n_i\textnormal{ odd }\\
\G_e&\textnormal{ Red and Black }&\textnormal{ Green and Blue }\\
\cline{1-3}
2&(0,0,S_z)&(S_x,S_y,0)\\
2'&(S_x,S_y,0)&(0,0,S_z)\\
1'&(0,0,0)&(S_x,S_y,S_z)\\
\end{tabular}
\end{center}
\end{table}

%-------------- Table 7 -------------------------------------------

\begin{table}
\caption{Restrictions on the form of $\sv(\kv)$ for any wave vector
  $\kv$ in the magnetic lattice when $\G_e=222$ or $22'2'$. In each
  case the form of $\sv(\kv)$ depends on the parities of $n_1 + n_3$ and 
  $n_2 + n_4$, where $\kv=\sum n_i\bi$. Colors refer to the points in
  Figure~\ref{fig:srplot}.}
\label{tab:sr2}
\setlength{\extrarowheight}{3pt}
\begin{center}
\begin{tabular}{>{$}c<{$}|>{$}c<{$}>{$}c<{$}>{$}c<{$}>{$}c<{$}}
%\hline
\cline{1-5}
 &n_1+n_3\textnormal{ even}&n_1+n_3\textnormal{ odd}
&n_1+n_3\textnormal{ odd}&n_1+n_3\textnormal{ even}\\ 
&n_2+n_4\textnormal{ even}&n_2+n_4\textnormal{ odd}
&n_2+n_4\textnormal{ even}&n_2+n_4\textnormal{ odd}\\
\G_e&\textnormal{ Red }&\textnormal{ Black }
&\textnormal{ Green }&\textnormal{ Blue }\\
\cline{1-5}
%\hline
 222&(0,0,0)&(0,0,S_z)&(S_x,0,0)&(0,S_y,0)\\
%\hline
 22'2'&(0,0,S_z)&(0,0,0)&(0,S_y,0)&(S_x,0,0)\\
%\hline
\end{tabular}
\end{center}
\end{table}

%-------------- Table 8 -------------------------------------------

%\onecolumn
\begin{table}
\caption{Additional restrictions on the form of $\sv(\kv)$ for special
  wave vectors that are invariant under mirror reflections when $G=8mm$.
  Note that in most cases the group of spin-space rotations $\G$ is
  abelian (except when $\G$ contains a 4-fold or an 8-fold rotation,
  with optional primes) in which case $\d^{i-1}\mu\d^{1-i}=\mu$, and
  $\d^i\mu\d^{1-i}=\d\mu$, and the selection rules take a much
  simpler form. Vectors $\kv_i$ along mirrors $m_i$ with $n_i$ odd,
  and vectors $\kv'_i$ along mirrors $m'_i$ with $n_i+l_i$ odd, are
  represented as open circles in Figure~\ref{fig:srplot}. }   
\label{tab:srmirrors}
\setlength{\extrarowheight}{3pt}
\begin{center}
\begin{tabular}{>{$}c<{$} | >{$}c<{$}| >{$}c<{$}}
\cline{1-3}
 \textnormal{Spin space-} & \kv_i=n_i\bi+l_i(\b{i-1}+\b{i+1})
 & \kv'_i=n_i(\bi+\b{i+1})+l_i(\b{i-1}+\b{i+2})\\
 \textnormal{group\ type} &  \text{ along } m_i 
 & \text{ along } m'_i\\
\cline{1-3}
P8^\d m^\mu m(\G_e)   & \d^{i-1}\mu\d^{1-i}\sv(\kv_i)=\sv(\kv_i)
                      & \d^i\mu\d^{1-i}\sv(\kv'_i)=\sv(\kv'_i)\\
\cline{1-3}
P8^\d b^\mu m(\G_e)   
 & \begin{cases}
     \d^{i-1}\mu\d^{1-i}\sv(\kv_i)=\sv(\kv_i) & \text{if $n_i$ even}\\
     \d^{i-1}\mu\d^{1-i}\sv(\kv_i)=-\sv(\kv_i) & \text{if $n_i$ odd}
   \end{cases}
 & \d^i\mu\d^{1-i}\sv(\kv'_i)=\sv(\kv'_i)\\
\cline{1-3}
P8^\d m_a^\mu m(\G_e) & \d^{i-1}\mu\d^{1-i}\sv(\kv_i)=\sv(\kv_i) 
                      
 & \begin{cases}
     \d^i\mu\d^{1-i}\sv(\kv'_i)=\sv(\kv'_i) & \text{if $n_i+l_i$ even}\\
     \d^i\mu\d^{1-i}\sv(\kv'_i)=-\sv(\kv'_i) & \text{if $n_i+l_i$ odd}
   \end{cases}\\
\cline{1-3}
P8^\d b_a^\mu m(\G_e) 
 & \begin{cases}
     \d^{i-1}\mu\d^{1-i}\sv(\kv_i)=\sv(\kv_i) & \text{if $n_i$ even}\\
     \d^{i-1}\mu\d^{1-i}\sv(\kv_i)=-\sv(\kv_i) & \text{if $n_i$ odd}
   \end{cases}
 & \begin{cases}
     \d^i\mu\d^{1-i}\sv(\kv'_i)=\sv(\kv'_i) & \text{if $n_i+l_i$ even}\\
     \d^i\mu\d^{1-i}\sv(\kv'_i)=-\sv(\kv'_i) & \text{if $n_i+l_i$ odd}
   \end{cases}\\
\cline{1-3}
\end{tabular}
\end{center}
\end{table}
%\twocolumn

%-------------- Figure 1 ------------------------------------------

\begin{figure}
  \centering
  \framebox{\framebox{Choice of $L$ from a known Bravais class}}\\
  $\downarrow$\\
  \framebox{Choice of $G$}\\
  $\downarrow$\\
  \framebox{Choice of $G_\eps$}\\
  $\downarrow$\\
  \framebox{Calculation of possible $\G$ and $\G_e$}\\
  $\downarrow$\\
  \framebox{Pairing of cosets}\\
  $\downarrow$\\
  \framebox{\framebox{Spin geometric crystal classes}}\\
  $\downarrow$\\
  \framebox{Choice of relative orientation of $L$ and $G$}\\
  $\downarrow$\\
  \framebox{\framebox{Spin arithmetic crystal classes}}\\
  $\downarrow$\\
  \framebox{Choice of gauge}\\
  $\downarrow$\\
  \framebox{Solution of group compatibility conditions}\\
  $\downarrow$\\
  \framebox{\framebox{Spin space-group types}}
  \caption{Flowchart describing the steps required for the enumeration
  of spin point groups and spin space-groups. Double boxes indicate
  the classification into equivalence classes as described in
  section~\ref{sec:classification}}
  \label{fig:flow}
\end{figure}

%-------------- Figure 2 ------------------------------------------

%-------------- Figure 3 ------------------------------------------

\begin{figure}
\caption{(Color) A subset of the wave vectors of the 2-dimensional octagonal
  lattice, exhibiting all possible selection rules.  The lattice
  generating vectors $\b{i}$ and their negatives, as well as the
  origin, are denoted by solid black circles. The rest of the points
  shown are of the form $\kv=n_1 \b1 + n_2 \b2 + n_3 \b3 + n_4 \b4$,
  with indices running from $-6$ to $6$.  Colors encode the parities
  of the indices of $\kv$ at each point as follows, Red: $n_1+n_3$ and
  $n_2+n_4$ both even; Black: $n_1+n_3$ and $n_2+n_4$ both odd; Blue:
  $n_1+n_3$ even and $n_2+n_4$ odd; Green: $n_1+n_3$ odd and $n_2+n_4$
  even. These color codes shold be used together with Tables
  \ref{tab:sr1} and \ref{tab:sr2} to determine the selection rules at
  each wave vector that are due to the lattice spin group $\G_e$.
  Vectors $\kv_i=n_i\bi+l_i(\b{i-1}+\b{i+1})$ invariant under mirrors
  $m_i$ with $n_i$ odd, and vectors
  $\kv'_i=n_i(\bi+\b{i+1})+l_i(\b{i-1}+\b{i+2})$ invariant under
  mirrors $m'_i$ with $n_i+l_i$ odd, are represented as open circles.
  These points should be used together with Table~\ref{tab:srmirrors}
  in determining the additional selection rules for wave vectors along
  mirror lines, when the point group is $8mm$.  }
\label{fig:srplot}
\resizebox{!}{18cm}{\includegraphics{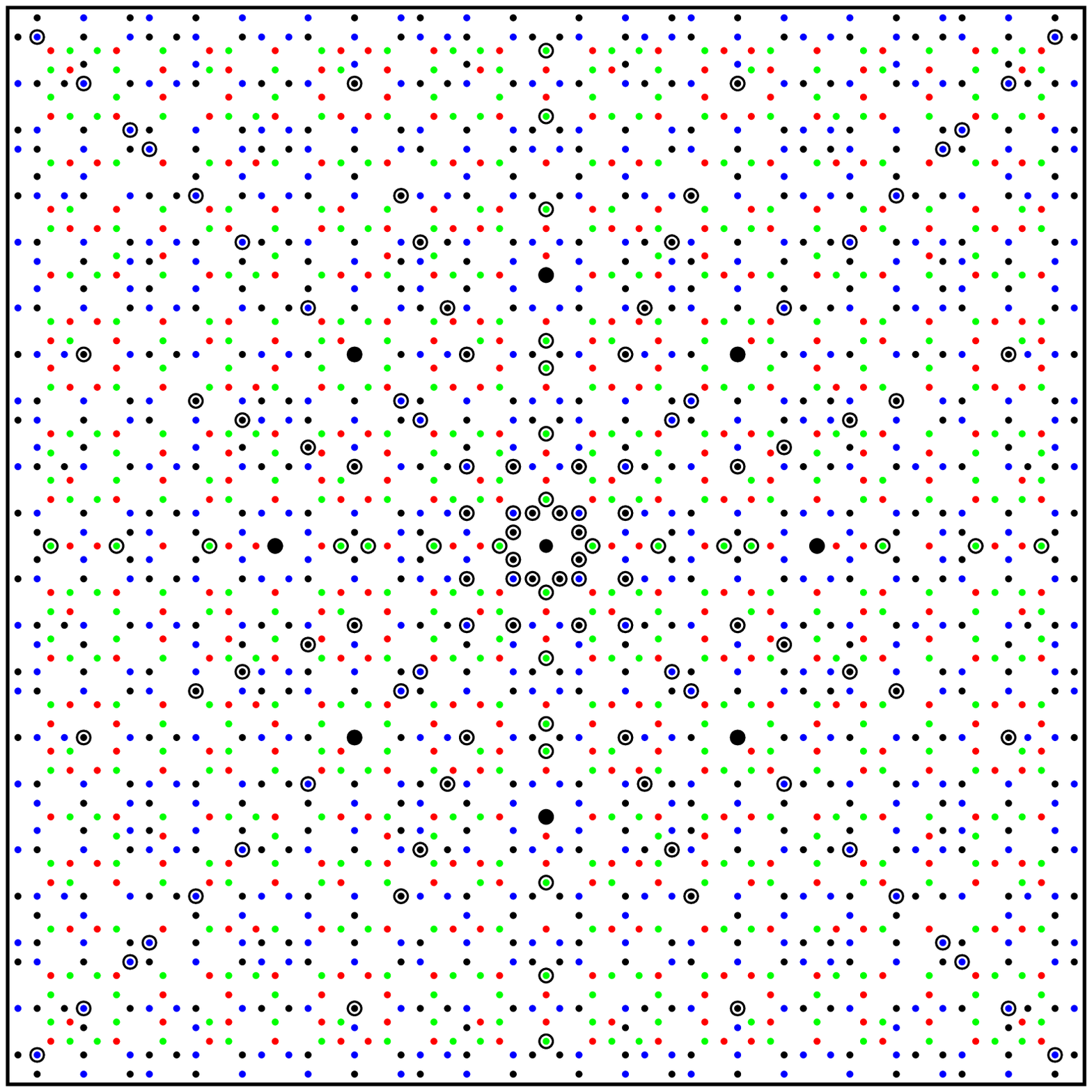}}
\end{figure}

\end{document}